\documentclass[10pt]{article}
\usepackage{amssymb}
\usepackage{amsmath}
\numberwithin{equation}{section}
\usepackage{graphicx}
\newtheorem{defi}{Definition}[section]
\newtheorem{rem}{Remark}[section]
\newtheorem{thm}{Theorem}[section]
\newtheorem{lemma}[thm]{Lemma}
\newtheorem{cor}[thm]{Corollary}
\newtheorem{prop}[thm]{Proposition}

\newtheorem{example}{Example}

\newcommand\qed{\hfill\blacksquare\bigskip}

\newcommand\mult{\mbox{mult}}
\newcommand\PP{\ensuremath{\mathbb{P}}}
\newcommand\ZZ{\ensuremath{\mathbb{Z}}}
\newcommand\CC{\ensuremath{\mathbb{C}}}

\newcommand\NN{\ensuremath{\mathbb{N}}}

\newcommand\vect[1]{\mbox{\boldmath$#1$}}
\newcommand\tee{\mathcal{T}}

\newcommand\zet[1]{\left\vert {#1} \right\vert}

\newcommand\proof{{\textbf {Proof.}}\ \,}
\begin{document}
\title{Solution of the generalized periodic discrete Toda equation}
\author{Shinsuke Iwao\\
Graduate School of Mathematical Sciences, \\
The University of Tokyo, \\
3-8-1
Komaba Meguro-ku, Tokyo \ 153-8914, Japan}
\date{}
\maketitle
\subsection*{Abstract}
A box-ball system with more than one kind of balls is obtained by the 
generalized periodic discrete Toda equation (pd Toda eq.).
We present an algebraic geometric study of the periodic Toda equation.
The time evolution of pd Toda eq. is linearized on the Picard group of
an algebraic variety,
and theta function solutions are obtained.
%\end{abstract}
%\pacs{02.30.Ik, 05.45.Yv}
%\ams{37K15, 37K20}
\section{Preface}\label{sec1}
A cellular automaton (CA) is a discrete dynamical system in which the
dependent variables take 
on a finite set of discrete values. Although CAs evolve in time by simple evolution
rules, they often show very complicated behaviour 
\cite{W}.

The box-ball system (BBS) is 
an important CA in which finitely many balls move in an array
of boxes under a certain evolution rule \cite{takahasi1,takahasi2}.
This discrete dynamical system is obtained from a discrete analogue
of the Toda equation through the limiting procedure ``ultradiscretization", and
displays the behavioural characteristics of nonlinear integrable
equations \cite{Hirota,Nagai}.
In fact, the BBS has soliton-like solutions and a large number of conserved
quantities \cite{MIT,MIT2}. 
Moreover, the solution of the initial value problem of the periodic
box-ball system (pBBS) with one kind of ball
has been obtained by ultradiscretizing the solution of the periodic discrete
Toda equation
(pd Toda).

In 1999, Tokihiro, Nagai and Satsuma  
pointed out that the pBBS with $M$ ($M\geq 1$)
kind of balls (and with capacity one)
is obtained by ultradiscretizing the generalized periodic discrete
Toda equation \cite{Tokihiro}.
%In \cite{MIT}, Mada, Idzumi and Tokihiro have established an algorithm to
%construct the conserved quantities of pBBSs by means of the ultradiscretization
%of the nonautonomas discrete KP equation.
%On the other hand,
Tokihiro and the author performed the
ultradiscretization of 
theta function solutions for $M=1$
of the pd Toda \cite{Iwao}.
It is to be expected that the solution of the initial value problem of the
pBBS with
$M$ kinds of balls can be obtained from the solution of the
generalized Toda equation,
as was the case for $M=1$. 

In this paper, we study the generalized pd Toda ($M\geq 1$):
\begin{align}
&I_n^{t+M}=I_n^t+V_n^t-V_{n-1}^{t+1},\label{toda1}\\
&V_n^{t+1}=\frac{I_{n+1}^tV_n^t}{I_n^{t+M}},\label{toda2}\\
&0<\textstyle\prod_{n=1}^N{V_n^t}<\textstyle \prod_{n=1}^N{I_n^t},\label{toda3}\\
&0<\textstyle\prod_{n=1}^N{V_n^t}<\textstyle \prod_{n=1}^N{I_n^{t+M-1}},\label{toda4}
\end{align}
with periodicity condition $I_{n+N}^t=I_n^t,V_{n+N}^t=V_n^t$,
where $N,n,t\in\NN$.

The time evolution of this system is linearized on the Picard group of
an
algebraic variety.
For example, the time evolution of pd Toda with $M=1$ is linearized
on Jacobi varieties of hyperelliptic curves \cite{Kimijima}.
For general $M$, the algebraic curves which appear in the linearization
are more complicated than hyperelliptic curves. 

These algebraic curves were analyzed by P. van Moerbeke and D. Mumford in 1978
\cite{Mumford}. In their work, they clarify the algebro-geometric properties
of the curves which are defined by so-called \textit{regular periodic operator} matrices.
Although our matrices are not always regular operators, their results are
essentially
also applicable to our case.

The special feature of the discrete system (\ref{toda1}-\ref{toda4})
is that we can explicitly
determine the action of
the unit time evolution $t\mapsto t+1$ on the Picard group
(proposition
\ref{time-shift}). Using this result, we extended the theta function expression
of the solution of pd Toda to the $M\geq 1$ case.

The paper is organized as follows. 
In section \ref{sec2},
we introduce a \textit{spectral curve} associated to the discrete system 
(\ref{toda1}-\ref{toda4})
and discuss its algebro-geometric
properties. Our aim in this section is
to determine the actions of the index shift $n\mapsto n+1$ and the
time shift $t\mapsto t+1$
on a Picard group $\mbox{Pic}^d(C)$ of this spectral curve $C$.
In section \ref{sec3},
we give the theta function expression (theorem \ref{seoremu})
which is the extension of the formula obtained by Kimijima and Tokihiro \cite{Kimijima}. 
\section{Spectral curve associated with pd Toda}\label{sec2}
\subsection{The nature of the spectral curve}
The pd Toda equation (\ref{toda1}), (\ref{toda2}) is equivalent to
the following matrix form:
\begin{equation}\label{matrixform}
L_{t+1}(y)R_{t+M}(y)=R_t(y)L_t(y),
\end{equation}
where $L_t(y)$ and $R_t(y)$ are given by
\[{}
L_t(y)=
\left(\begin{array}{@{\,}cccc@{\,}}
	1 &  & & V_N^t\cdot 1/y \\
	V_1^t & 1 & &  \\
	 & \ddots & \ddots & \vdots \\
	 &  & V_{N-1}^t & 1 
\end{array}\right),\quad
R_t(y)=
\left(\begin{array}{@{\,}cccc@{\,}}
	I_1^t & 1 &  &  \\
	 & I_2^t & \ddots &  \\
	 &  & \ddots & 1 \\
	y &  &  & I_{N}^t 
\end{array}\right),
\]
and $y$ is a complex variable. Let us introduce a new matrix
$X_t(y)$ defined by
\begin{equation}\label{ex}
X_t(y):=L_t(y)R_{t+M-1}(y)\cdots R_{t+1}(y)R_t(y).
\end{equation}
From (\ref{matrixform}) and (\ref{ex}), we obtain
\begin{equation}\label{Lax}
X_{t+1}(y)R_{t}(y)=R_t(y)X_t(y),
\end{equation}
which implies that the eigenvalues of $X_t(y)$ are conserved quantities under the time
evolution. 
\begin{lemma}\label{new-lemma}
Assume that $\{I_n^t,V_n^t\}$ satisfies pd Toda equation 
$\mathrm{(\ref{toda1}\mbox{--}\ref{toda4})}$.
Then,
$\prod_{n=1}^N{V^t_n}=\prod_{n=1}^N{V^{t+1}_n}$ and
$\prod_{n=1}^N{I^t_n}=\prod_{n=1}^N{I^{t+M}_n}$.
\end{lemma}
\proof
By (\ref{Lax}),
$\det{X_t(y)}\!\!=\!\!y^{-1}\!\left(y-\prod_n{V_n^t}\right)\left(\prod_n{I_n^t}-y\right)
\!\cdots\!\left(\prod_n{I_n^{t+M-1}}\!-y\right)$ 
does not change under the time evolution for any $y$.
Then, 
the set of quantities $\mathcal{U}_t:=
\{\prod_n{V_n^t},\prod_n{I_n^t},\dots,\prod_n{I_n^{t+M-1}}\}$
satisfies $\mathcal{U}_t=\mathcal{U}_{t+1}$.
From this equation, it follows that $\{\prod_n{V_n^t}
,\prod_n{I_n^t}\}=\{\prod_n{V_n^{t+1}},\prod_n{I_n^{t+M}}\}$.
The inequality (\ref{toda3}) and (\ref{toda4}) leads the lemma.
$\qed$

Let $\widetilde{\Phi}(x,y):=\det{(X_t(y)-xE)}$ be the characteristic
polynomial of $X_t(y)$ ($E$ is the unit matrix).
%Let us define the algebraic curve $\widetilde{C}$ by the algebraic equation
The equation
\begin{equation}\label{Phi}
\widetilde{\Phi}(x,y)=0
\end{equation}
defines the affine part $\widetilde{C}$ of its completion $C$.
We call this projective curve $C$ the
\textit{spectral curve} $C$ of the pd Toda equation.
%A \textit{spectral curve} $C$ of the pd Toda equation is the projective curve defined 
%as the completion of $\widetilde{C}$.
$C$ is a $(M+1)$-fold ramified covering over $\PP^1$:
\begin{equation}\label{ekkusu}
p_x:C\to\PP^1,
\end{equation}
and is also a $N$-fold ramified covering
$p_y:C\to\PP^1$.
It goes without saying that $C$ is conserved under the time
evolution and 
is completely determined
by the initial values
$\{V_n^0,I_n^0,I_n^1,\dots,I_n^{M-1}\}_{n=1}^N$.
Note that $C$ may fail to be smooth in certain situations.
We restrict ourselves to the case where $C$ is smooth.
%\begin{lemma}\label{smooth}
\begin{rem}
If $M<N$,
the matrix $X_t(y)$ is a $N\times N$ matrix and is of the form
\begin{align}
X(y)&=\left(\begin{array}{@{\,}ccccccccccc@{\,}}
	\alpha^{(1)}_1 & \alpha^{(2)}_2 & \cdots & \alpha^{(M)}_M & 1         & 0 & & &\\%[-1pt]
\beta_1 &  \alpha^{(1)}_2& \alpha^{(2)}_3 & \cdots & \alpha^{(M)}_{M+1} & 1 &   0&  &   \\%[-1pt]
0       & \beta_2 &\alpha^{(1)}_3 &   \alpha^{(2)}_4&\cdots&\alpha^{(M)}_{M+2} &1 &0&\\%[-4pt]
	 & 0      & \ddots  & \ddots  & \ddots & & \ddots&\ddots  \\%[-4pt]
	 &         &        &   &  & &        &    &1 \\
	 &          &      &    & &  &     &        &\alpha^{(M)}_N\\
           &     &        &     &  & &  \ddots   &   \ddots     &  \vdots \\
           &     &        &    &   &  & 0   &  \beta_{N-1}    &  \alpha^{(1)}_N
\end{array}\right)\nonumber
\end{align}
\begin{align}
&+\left(\begin{array}{@{\,}ccccccc@{\,}}
	0 &  &  &  &    \\
	\vdots & &  &  &    \\[-1mm]
        1 & & & & & \\
	\alpha^{(M)}_1 &\ddots  &  &     \\
        \vdots &   \ddots    &   &  &       \\
	\alpha^{(3)}_1 & & \ddots & \ddots    \\
	\alpha^{(2)}_1 & \alpha^{(3)}_2 &\cdots& \alpha^{(M)}_{M-1} & 1&\cdots&0    
\end{array}\right)\!\times\! y
+\left(\begin{array}{@{\,}ccc@{\,}}
	 &  & \beta_N \\
	 &  &  \\
	 &  & 
\end{array}\right)\!\times\! \frac{1}{y},\label{matrix}
\end{align}
where $\alpha_j^{(i)},\beta_j,(1 \leq i\leq M,1\leq j\leq N)$ are
polynomials in $\{V_n^t,I_n^t,I_n^{t+1},\dots,$
$I_n^{t+M-1}\}_{n=1}^N$.
In general, the $(i,j)$-component of $X(y)$
is the essentially finite summation
$(X(y))_{i,j}=\sum_{l=-1}^\infty{\alpha_j^{(j-i+lM+1)}y^l}$,
where $\alpha_j^{(-1)}=\beta_j$, $\alpha_j^{(M+1)}=1$ and $\alpha_j^{(P)}=0$
$(P<-1,P>M+1)$.
\end{rem}
\begin{rem}
When the greatest common divisor $(N,M)\neq 1$,
this particular matrix $(\mathrm{\ref{matrix}})$ is not a regular periodic
difference operator which is analyzed in $\mathrm{\cite{Mumford}}$.
\end{rem}
%\begin{lemma}\label{smooth}
%$\widetilde{C}$ is smooth for generic initial values.
%\end{lemma}
%\proof
%It is enough to prove that
%\[
%\mbox{Dis}_x(\mbox{Dis}_y\Phi(x,y))\in\ZZ\,[\{V_n^0,I_n^0,I_n^1,\dots,I_n^{M-1}\}_{n=1}^N]
%\]
%is a non zero element of $\ZZ\,[\{V_n^0,I_n^0,I_n^1,\dots,I_n^{M-1}\}_{n=1}^N]$,
%where $\mbox{Dis}_XF(X,Y,\cdots)$ is a discriminant of $F$ as 
%a polynomial in $X$.
%It is easy to check this. For example, substitute
%$I_n^0=I_n^1=\dots=I_n^{M-1}=0$ for all $1\leq n\leq N$,
%and calculate $\mbox{Dis}_x(\mbox{Dis}_y\Phi(x,y))\in\ZZ[V_1^0,\dots,V_N^0]$.
%It is easy to prove:
%\[
%\overline{F}\neq 0\in\ZZ[V_1^0,\dots,V_N^0]\Rightarrow
%F\neq 0\in\ZZ\,[\{V_n^0,I_n^0,I_n^1,\dots,I_n^{M-1}\}_{n=1}^N],
%\]
%where $\overline{F}=F\vert_{I_n^0=I_n^1=\dots=I_n^{M-1}=0\,,\,n=1,\dots,N}$.$\qed$
%\begin{rem}
%Although lemma \ref{smooth} seems to be obvious
%in our case, some integrable
%systems do not satisfy this property.
%\end{rem}

%Let $\widetilde{C}$ be smooth curve defined by (\ref{Phi})
%and
%$C\supset\widetilde{C}$ be the completion.

We now analyze
the points contained in $C\setminus\widetilde{C}$.
We calculate the polynomial expression of the function
$\widetilde{\Phi}(x,y)$ in $x$, $y$ and
$y^{-1}$:
\begin{equation}\label{expression}
{} 
\widetilde{\Phi}(x,y)=A_0(x)y^{M}+A_1(x)y^{M-1}+\dots+A_M(x)+
A_{M+1}(x)y^{-1}=0.
\end{equation}
For the analysis of the behaviour of $C$,
we need to analyze $A_j(x)\,,(j=0,1,\dots,$ $M+1)$.
The following lemma is established by van Moerbeke and Mumford \cite{Mumford}.
They analyzed the determinant $\det{(X(y)-xE)}$ by direct calculation.
\begin{lemma}\label{mum}
Let $m:=(N,M)$, $N=mN_1$ and $M=mM_1$.
The polynomial $A_j(x)$ is a polynomial of degree $k_j$ satisfying
\begin{equation}\label{degree}
k_j\leq \frac{jN}{M},\ (0\leq j\leq M), \quad k_{M+1}=0.
\end{equation}
The equality in (\ref{degree}) holds if and only if the right hand side is an integer.
Moreover, $A_0(x),A_{M_1}(x),A_{2M_1}(x),\dots,A_{mM_1}(x)$ are expressed as:
\[
A_{rM_1}(x)=(-1)^{M(N-M)+r}\left(\begin{array}{@{\,}c@{\,}}
	m \\
	r
\end{array}\right)
x^{rN_1}+\cdots,\quad(r=0,1,\dots,m).
\]
\end{lemma}
We start form the polynomial (\ref{expression}).
Let $\gamma:=x^{N_1}y^{-M_1}$.
By lemma \ref{mum}, we obtain the expression
\begin{eqnarray*}
{}
y^{-mM_1}\widetilde{\Phi}(x,y)&=(-1)^{M(N-M)}\left(\gamma^m
-{m \choose 1}\gamma^{m-1}+{m \choose 2}\gamma^{m-2}+\dots
+(-1)^m\right)\\
&\hspace{3cm}+\mbox{ lower order terms when } \zet{x},\zet{y}\to\infty \\
&\sim (-1)^{M(N-M)}(\gamma-1)^m.
\end{eqnarray*}
This implies $x^{N_1}y^{-M_1}=\gamma\sim 1$ near $(x,y)=(\infty,\infty)$.
By $(N_1,M_1)=1$, there exists the local coordinate $t$
equipped
with the completion $C\supset\widetilde{C}$
such that 
\[
x\sim t^{-mM_1}=t^{-M}
,\ y\sim t^{-mN_1}=t^{-N}
,\quad (x,y)\sim(\infty,\infty).
\]
In particular, there exists \textit{only one} point $P\in C$ which is expressed as
$P=(\infty,\infty)$.

In a similar manner, there exists a point $Q\in C$ which is expressed as
$Q=(\infty,0)$. The local coordinate $t$ near $Q$ satisfies
$
x\sim t^{-1},\ y\sim t^{N}.
$
Using these fact, the divisors $(x),(y)\in\mbox{Div}(C)$ are
\begin{align}
(x)&=-MP-Q+(\mbox{a positive divisor on $\widetilde{C}$}),\\
(y)&=-NP+NQ.
\end{align}
\begin{rem}
The existence of the unique point $P(\infty,\infty)$
is a special property of $X(y)$. In fact, there exist
$m$ points $P_j(x,y)=(\infty,\infty)$ $(j=1,\dots,m)$ on the algebraic curve
associated to a regular operator matrix \cite{Mumford}.
\end{rem}
Although the concrete calculations in \cite{Mumford}
should be applied only to the case that $X(y)$ is a regular operator,
these results are also applicable to our case
on condition that $C$ is smooth.
Precisely, these calculations 
become true for our case by substituting
\begin{equation}
P_1=P_2=\dots=P_m(=P).
\end{equation}

\subsection{The eigenvector mapping}
We now define the \textit{isolevel set} $\tee_C$ as the set of matrices $X(y)$
associated with the
the spectral curve $C$. The following proposition is fundamental
to the algebro-geometric method for integrable systems.
\begin{prop}\label{bundle}
Let $C$ be smooth and $X(y)\in\tee_C$.
There is a unique line bundle
$V\subset C\times \CC^{N}$
s.t. 
\[{}
\pi^{-1}(x,y)=\{\mbox{the eigenspace of $X(y)$
corresponding to the eigenvalue $x$}
\}\subset\CC^{N},
\]
for all $(x,y)\in C\setminus\{x=0,\infty\}$
where
$
C\times \CC^{N}\supset V \stackrel{\pi}{\longrightarrow} C
$
is called the canonical projection.
\end{prop}

By virtue of this proposition,
we obtain the map
\[
\begin{array}{cccc}
 \varphi_C:&\tee_C & \to&\{U\to\PP^{N-1}
\}_{U\subset C}\\[1mm]
 &X(y) &\mapsto & V^\lor
\end{array},
\]
where $V^\lor$ is the dual bundle of $V$.
The section of $V^\lor$ is a component of
%$\{(x,y)\mapsto \vect{v}(x,y)\}_{(x,y)\in C}$
%where $\vect{v}(x,y)\in \PP^{N-1}$ is 
the eigenvector of $X(y)$.
By the Grothendieck-Riemann-Roch theorem, it follows that
\begin{equation}\label{GRR}
\mbox{Im}\,\varphi_C\subset\mbox{Pic}^d(C),\quad d=g+N-1,
\end{equation}
where $g$ is the genus of $C$. (See \cite{Top}).
\begin{defi}
For smooth $C$, the eigenvector mapping associated to the equation
$(\mathrm{\ref{Lax}})$ is the mapping
\[
\varphi_C:\tee_C\to\mbox{Pic}^d(C)
\]
defined as above.
We shall call $V^\lor$ the eigenvector bundle.
\end{defi}

The eigenvector mapping is an important tool to analyze
the various integrable systems \cite{Griffith}. The following proposition is essential
to our arguments in the present paper.
\begin{prop}
The eigenvector mapping 
$
\varphi_C:\tee_C\to\mbox{Pic}^d(C)
$
is an isomorphism to $\mathrm{Im}\varphi_C$.
\end{prop}
This proposition is a straightforward result of the following theorem
provided by van Moerbeke and Mumford.
\begin{thm}\label{oneone}
There is a one-to-one correspondence between the two sets of data:\\
$(\mathrm{a})$a multi diagonal matrix $X(y)$ of the form (\ref{matrix}) such
that $\widetilde\Phi(x,y)=0$ defines an affine part of
a smooth curve.\\
$(\mathrm{b})$a smooth curve $C$, two points $P,Q$, 
 two functions $x,y$ on $C$
and a divisor $\mathcal{D}$ which satisfies
\begin{equation}\label{jyuuyou}
\varphi_C(X(y))=\mathcal{D}+(N-1)Q.
\end{equation}
$C$ has genus
$
\displaystyle g=\frac{(N-1)(M+1)-m+1}{2},
$
and $\deg{\mathcal{D}}=g$.
\end{thm}
\begin{rem}\label{kako}
Although
the equation $(\mathrm{\ref{jyuuyou}})$ does not appear
in the van Moerbeke and Mumfords paper \cite{Mumford},%refa
we easily derive this equation from the relation
$($p.107$)$%ref
\begin{equation}\label{arigato}
(g_k)+\mathcal{D}\geq \sum_{i=k+1}^N{P_i}-\sum_{i=k}^{N-1}{Q_i},\quad k=1,2,\dots,N-1,
\end{equation}
where $(g_1,\dots,g_{N-1},1)^T$ is a section of the eigenvector bundle $V^\lor$.
In fact, $(\mathrm{\ref{arigato}})$ yields 
\begin{equation}\label{futousiki}
(g_k)_\infty\leq \mathcal{D}+(N-k)Q,
\end{equation}
which implies $d=\deg{(\mathrm{Im}\varphi_C)}\leq \deg{\mathcal{D}}+N-1=g+N-1$.
Because of the equality $(\mathrm{\ref{GRR}})$, $(g_1)_\infty$ must satisfy
\begin{equation}\label{tousiki}
(g_1)_\infty=\mathcal{D}+(N-1)Q.
\end{equation}
$(\mathrm{\ref{futousiki}})$ and $(\mathrm{\ref{tousiki}})$ imply
$(\mathrm{\ref{jyuuyou}})$.
\end{rem}
\begin{defi}
A finite component $\varphi_{\mathrm{fn}}(X(y))$
of the eigenvector mapping is a positive divisor $\mathcal{D}$ of
degree $g$ which appears in $(\mathrm{\ref{jyuuyou}})$.
\end{defi}
%The divisor
%$\varphi_{\mathrm{fn}}(X(y))$ satisfies 
%$
%\varphi(X(y))=\varphi_{\mathrm{fn}}(X(y))+(N-1)Q
%$
%by definition.
\subsection{The action of the evolutions on the eigenvector bundle}\label{two-two}
In this section, we represent the two actions ---
index evolution $(n\mapsto n+1)$ and time evolution $(t\mapsto t+1)$ --- on the eigenvector
bundle.
\begin{prop}\label{index-shift}
Let $D_n$ be the divisor
$
D_n=P-Q.
$
Then the following 
diagram is commutative.
\[
\begin{array}{ccccc}
 &\tee_C& \to &\mbox{Pic}^d(C)& \\[1mm]
{}_{n\mapsto n+1}& \downarrow& &\downarrow&\hspace{-4mm}{}_{+D_n} \\[1mm]
  & \tee_C& \to& \mbox{Pic}^d(C)&
\end{array}
\]
\end{prop}
\proof 
Let us denote $\sigma:n\mapsto n+1$. A straightforward calculation
leads to
\begin{equation}\label{guiter}{}
X(y)\left(\begin{array}{@{\,}c@{\,}}
	v_1 \\
	\vdots \\
	v_{N-1} \\
	v_N
\end{array}\right)
=x\left(\begin{array}{@{\,}c@{\,}}
	v_1 \\
	\vdots \\
	v_{N-1} \\
	v_N
\end{array}\right)
\
\Leftrightarrow \ 
\sigma^{-1}(X(y))\left(\begin{array}{@{\,}c@{\,}}
	y^{-1}v_{N} \\
	 v_1\\
	\vdots \\
	v_{N-1}
\end{array}\right)
=x\left(\begin{array}{@{\,}c@{\,}}
	y^{-1}v_{N} \\
	 v_1\\
	\vdots \\
	v_{N-1}
\end{array}\right).
\end{equation}
It is enough to prove
%\begin{equation}
$(y^{-1}g_{N-1}^{-1})_\infty-(g_1)_\infty\sim -D_n$.
%\end{equation}
By (\ref{arigato}), we have
%\begin{equation}
$
(y\,g_{N-1})+\mathcal{D}\geq
%\sum_{j=0}^{N-2}{Q_j}-\sum_{j=1}^{N-1}{P_j}.
(N-1)Q-(N-1)P.
$
%\end{equation}
An argument similar
to remark \ref{kako} allows to conclude
$
(y^{-1}g_{N-1}^{-1})_\infty\sim(y\,g_{N-1})_\infty=\mathcal{D}+(N-1)P,
$
which completes the proof.$\qed$
%Let $\mathcal{F}$ be an invertible sheaf associated with $V^\lor$.
%Suppose the direct image \[(p_x)_\ast\mathcal{F},\]
%that is a sheaf on $\PP^1$.
%By definition,
%$(p_x)_\ast\mathcal{F}$ has
%at 
%least $N$ independent sections
%\[
%(p_x)_\ast v_1,\,(p_x)_\ast v_2\,,\cdots, (p_x)_\ast v_N,
%\]
%which $v_j(x,y)$ is a $j$-th component of the eigenvector of $X(y)$.
%
%Let $V\stackrel{\pi}{\to}C$ be the line bundle in proposition \ref{bundle}.
%Suppose the map $f:=x\circ \pi$:
%\begin{equation}\label{projection}
%V\stackrel{\pi}{\to}C \stackrel{x}{\to} \PP^1\ (x\mbox{-plane}).
%\end{equation}

In order to determine the action of the
time evolution on the eigenvector bundle,
we introduce the concepts of
\textit{Bloch solution} and \textit{transposed operator}.
%we need to prove the following proposition. 
%\begin{prop}\label{time-shift}
%Let $D_t$ be the divisor 
%$
%D_t=A-Q,
%$ 
%where $A=(0,y_0)$ 
%and $y_0$ is the complex number which satisfies $\det{R_t(y_0)}=0$.
%Then, the following 
%diagram is commutative.
%\[
%\begin{array}{ccccc}
% &\tee_C& \to &\mbox{Pic}^d(C)& \\[1mm]
%{}_{t\mapsto t+1}& \downarrow& &\downarrow&\hspace{-4mm}{}_{+D_t} \\[1mm]
%  & \tee_C& \to& \mbox{Pic}^d(C)&
%\end{array}
%\]
%\end{prop}

%To prove this proposition, we introduce the concepts of
%\textit{Bloch solution} and \textit{transposed operator}.

We identify the eigenvectors of $X(y)$
with the
\textit{Bloch solutions with multiplicity} $y$ of the periodic infinite matrix
\[{}
\widetilde{X}
=\left(\begin{array}{@{\,}ccccccccccc@{\,}}
 \ddots  &   \ddots    &  &\ddots & \ddots& & & & & \\
\ddots &\alpha^{(1)}_N & \alpha^{(2)}_1 & \cdots & \alpha^{(M)}_{M-1} & 1         & 0 & &\\
 & \beta_N &  \alpha^{(1)}_1& \alpha^{(2)}_2 & \cdots & \alpha^{(M)}_M & 1 &   0&     \\
 &0       & \beta_1 & \alpha^{(1)}_2 &   \alpha^{(2)}_3 &\cdots  &  \alpha^{(M)}_{M+1} &1 &0\\
 & & 0      & \ddots    &\ddots&  & & \ddots&\ddots  \\
 &	 &         &        &   &  & &        &   \\
\end{array}\right),
\]
which are the infinite vectors $\widetilde{\vect{v}}=
(\cdots,v_{n-1},v_n,v_{n+1},\cdots)^T$ such that
\begin{equation}\label{Bloch}
\widetilde{X}\widetilde{\vect{v}}=x\widetilde{\vect{v}}\quad
\mbox{ and }\quad v_{n+N}=y\,v_n.
\end{equation}
%and
%\begin{equation}
%v_{n+N}=y\,v_n.
%\end{equation}
The first equation of (\ref{Bloch}) is equivalent to
\begin{equation}\label{linear}
{}
\beta_{n-1}v_{n-1}+\alpha^{(1)}_nv_n+\alpha^{(2)}_{n+1}v_{n+1}+\cdots
+\alpha^{(M)}_{n+M-1}v_{n+M-1}+v_{n+M}=xv_n
\end{equation}
for $n\in\ZZ$\quad $(\alpha^{(j)}_{n+N}=\alpha^{(j)}_n,\,\beta_{n+N}=\beta_n)$.
%For fixed $x\in\PP^1$, $f^{-1}(x)\subset V$ is an union of the vector subspaces of
%$\CC^N$. It is also natural that we identify 
%\[
%\mbox{Span}_{\CC^N}\langle f^{-1}(x)\rangle
%\!=\!\{\mbox{the vector space of the Bloch solutions of 
%$\widetilde{X}$ belongs to $x$}
%\}.
%\]
%
%The mapping \[x\mapsto \mbox{Span}_{\CC^N}\langle f^{-1}(x)\rangle \quad
%\subset G(M,N),\]
%determines a family of holomorphic mappings
%\[
%\Phi:\PP^1\to G(M,N).
%\]
%The equation (\ref{Bloch}) is equivalent to
Because the l.h.s.\relax\ of
(\ref{linear}) is a linear combination of
$
v_{n-1},v_n,\cdots,v_{n+M-1}
$,
the Bloch solution associated with $(x,y)\in C$
are also be written as
a linear combination of
$
\widetilde{\vect{v}}^{(1)},\widetilde{\vect{v}}^{(2)},\dots,
\widetilde{\vect{v}}^{(M+1)}
$,
where 
\begin{eqnarray*}
\widetilde{\vect{v}}^{(1)}&=(\dots,1,0,\dots,0,v^{(1)}_{M+2},
v^{(1)}_{M+3},\dots)^T\\
\widetilde{\vect{v}}^{(2)}&=(\dots,0,1,\dots,0,v^{(2)}_{M+2},
v^{(2)}_{M+3},\dots)^T\\ [-0.3cm]
&\vdots \\
\widetilde{\vect{v}}^{(M+1)}&=(\dots,0,0,\dots,1,v^{(M+1)}_{M+2},
v^{(M+1)}_{M+3},\dots)^T.
\end{eqnarray*}
More
precisely, let $\psi(x,y)$ be the Bloch solution:
\begin{equation}\label{tenkai}
\psi(x,y)=a_1\widetilde{\vect{v}}^{(1)}+a_2\widetilde{\vect{v}}^{(2)}
+\dots+a_{M+1}\widetilde{\vect{v}}^{(M+1)},
\end{equation}
where
%\[
$a_i=a_i(x,y)$ % \quad
and $\widetilde{\vect{v}}^{(i)}=\widetilde{\vect{v}}^{(i)}(x)$.
%\] 

Recalling (\ref{Lax}) and proposition \ref{bundle}
the eigenvector $\vect{v}_t(x,y)$ at time $t$
satisfies
\begin{equation}\label{tuketasi}
\vect{v}_{t+1}(x,y)=R_t(y)\cdot\vect{v}_t(x,y).
\end{equation}
Equivalently, the Bloch solution $\psi_t(x,y)$ satisfies
\begin{equation}\label{hatten}
\psi^{t+1}(x,y)=\widetilde{R}_t\cdot\psi^t(x,y),
\end{equation}
where
\[
\widetilde{R}_t=
\left(\begin{array}{@{\,}ccccccc@{\,}}
   \ddots &\ddots     &  &  &  & &  \\
    &I_N^t&1& & & & \\    
     &	&I_1^t & 1 & & &  \\
    &	& & I_2^t & \ddots & & \\
     &	& &  & \ddots & 1 & \\
    &	& &  &  & I_{N}^t &\ddots \\
    &   &  & &  &   &\ddots
\end{array}\right).
\]
(\ref{tenkai}) and (\ref{hatten}) yield
\begin{equation}\label{a-hatten}
{}
a_j^{t+1}=a_1^t(\widetilde{R}_t\widetilde{\vect{v}}^{(1)})_j+\dots
+a_{M+1}^t(\widetilde{R}_t\widetilde{\vect{v}}^{(M+1)})_j,
\quad j=1,2,\dots,M+1,
\end{equation}
where $(\vect{v})_j$ is a $j$th-component of the vector $\vect{v}$.
Equation (\ref{a-hatten}) is equivalent to
\[
\left(a_1^{t+1},a_2^{t+1},\dots,a_M^{t+1},a_{M+1}^{t+1}\right)^T
=H_t\cdot\left(a_1^{t},a_2^{t},\dots,a_M^{t},a_{M+1}^{t}\right)^T,
\]
%\begin{equation}
%\left(\begin{array}{@{\,}c@{\,}}
%	a_1^{t+1} \\
%	a_2^{t+1} \\
%	\vdots \\
%	a_{M}^{t+1} \\
%	a_{M+1}^{t+1}
%\end{array}\right)
%=
%H_t\cdot\left(\begin{array}{@{\,}c@{\,}}
%	a_1^{t} \\
%	a_2^{t} \\
%	\vdots \\
%	a_{M}^{t} \\
%	a_{M+1}^{t}
%\end{array}\right),
%\end{equation}
where 
\begin{equation}\label{eiti}
H_t=
\left(\begin{array}{@{\,}ccccc@{\,}}
	I_1^t & 1 &  &  &  \\
	& I_2^t & \ddots &  &  \\
	 &  & \ddots & 1 &  \\
	 &  &  & I_M^{t} & 1 \\
	v_{M+2}^{(1)} & v_{M+2}^{(2)} & \cdots & v_{M+2}^{(M)} & I_{M+1}^t+v_{M+2}^{(M+1)}
\end{array}\right).
\end{equation}
Using this equation, we obtain
\begin{equation}\label{timeact}
\left(v_1^{t+1},v_2^{t+1},\dots,v_M^{t+1},v_{M+1}^{t+1}\right)^T
=H_t\cdot
\left(v_1^{t},v_2^{t},\dots,v_M^{t},v_{M+1}^{t}\right)^T,
\end{equation}
%\begin{equation}\label{timeact}
%\left(\begin{array}{@{\,}c@{\,}}
%	v_1^{t+1} \\
%	v_2^{t+1} \\
%	\vdots \\
%	v_{M}^{t+1} \\
%	v_{M+1}^{t+1}
%\end{array}\right)
%=H_t\cdot
%\left(\begin{array}{@{\,}c@{\,}}
%	v_1^{t} \\
%	v_2^{t} \\
%	\vdots \\
%	v_{M}^{t} \\
%	v_{M+1}^{t}
%\end{array}\right),
%\end{equation}
where 
$
\psi^t(x,y)=(\dots,v_1^t,v_2^t,\dots,v_{M+1}^t,\dots)^T
$.
\begin{lemma}\label{independent}
For fixed generic $x$,
the $M+1$ Bloch solutions associated with $x$ are linearly independent.
\end{lemma}
\proof
For generic $x$, associated multiplicities
$y_j$ $(j=1,2,\dots,M+1)$ of the Bloch solutions
are all distinct.
$\blacksquare$
\begin{lemma}\label{det}
$ \det{H_t}=(-1)^{M+1}I_1^tx.
$
\end{lemma}
This lemma is proved by an elementary calculation, which we shall give
in the appendix.

Let $\mathcal{F}$ be the invertible sheaf associated with the eigenvector bundle
$V^\lor$.
Let us consider the direct image
$
(p_x)_\ast\mathcal{F}
$,
where
$p_x$ is a projection of $C$ defined by (\ref{ekkusu}).
The sheaf $(p_x)_\ast\mathcal{F}$
is a locally free sheaf of rank $M+1$.
By lemma \ref{independent}, the fiber of this direct sheaf is $\CC^{M+1}$.

Equation (\ref{timeact})
is regarded as the time action to the space $\CC^{M+1}$.
%The germ of $(p_x)_\ast\mathcal{F}$ at $x\in\PP^1$ is of the form
%\[
%x\mapsto \{
%(p_x)_\ast v(x,y_1),\dots,(p_x)_\ast v(x,y_{M+1})
%\},
%\]
%where  $v(x,y)$ is a section of $\mathcal{F}$ and
%$\{(x,y_1),\dots,(x,y_{M+1})\}=p_x^{-1}(x)$.
%Let us consider
%\[
%(p_x)_\ast v(x,y_j)
%\]
%as a section
%\[
%x\mapsto (p_x)_\ast v(x,y_j).
%\]
Recall that the components of the Bloch solutions of $\widetilde{X}$
are the section of the sheaf $\mathcal{F}$. 
For $x\in\PP^1$, we denote the $k$-th component of the infinite vector $\psi^t(x,y_j)$
by $v_{jk}^t(x)$, and $\psi^t(x,y_j)$ by $\psi^t_j(x)$.
By (\ref{tenkai}), the finite vectors
$
\widehat{\psi}^t_j:=(v_{j,1}^t,v_{j,2}^t,\dots,v_{j,M+1}^t)^T
$
have the property
\begin{equation}\label{finite}
{}
\{\psi^t_j\}_{j=1}^{M+1}\mbox{ are linearly independent}
\Longleftrightarrow
\{\widehat{\psi}^t_j\}_{j=1}^{M+1}\mbox{ are linearly independent}.
\end{equation}
Equation (\ref{timeact}) implies that
\begin{equation}
(\widehat{\psi}^{t+1}_1,\widehat{\psi}^{t+1}_2,\dots,
\widehat{\psi}^{t+1}_{M+1})=H_t\cdot
(\widehat{\psi}^{t}_1,\widehat{\psi}^{t}_2,\dots,
\widehat{\psi}^{t}_{M+1}).
\end{equation}
On the other hand,
lemma \ref{independent}, \ref{det} and (\ref{finite}) imply
that there exists at least one vector $\widehat{\psi}^{t}_j$  which
satisfies
\begin{equation}\label{ineq}
\mult_x\widehat{\psi}^{t+1}_j>
\mult_x\widehat{\psi}^{t}_j,
\end{equation}
where
$\mult_x(z_1,z_2,\dots,z_N)^T=\min{
\left[\ \mult_xz_1,\mult_xz_2,\dots,\mult_xz_N
\ \right]}$ and \\
$\mult_x{z}$ is multiplicity of $x$ in $z$.

The same discussion can be repeated
for the projection
\[
p_y:C\to\PP^1,
\]
which is an $N$-fold ramified covering over $\PP^1$.

The following two facts are then obvious to prove.
\begin{lemma}\label{indep-y}
For fixed generic $y$,
the $N$ eigenvectors of $X_t(y)$ are linearly independent.
\end{lemma}
\begin{lemma}\label{det-y}
$\det{R_t(y)}$ is a polynomial of degree one
in $y$.
\end{lemma}

Let us consider the direct image
$
(p_y)_\ast\mathcal{F}
$.
From
lemma \ref{indep-y} we then find that the fiber of this direct image
is $\CC^N$.

Equation (\ref{tuketasi}) can be regarded as the time action
to the space $\CC^N$.
%By (\ref{ineq}) and lemma \ref{indep-y}, \ref{det-y},
For fixed $y$, we denote $p_y^{-1}(x)=\{(x_1,y),\dots,(x_{N},y)\}$.
Then we obtain that
\begin{equation}\label{nyan}
\exists \,j \quad\mbox{\ s.t.\ }\quad \mult_{(y-y_t)}\vect{v}_{t+1}(x_j,y)
>\mult_{(y-y_t)}\vect{v}_t(x_j,y),
\end{equation}
where $\det{R_t(y_t)}=0$
$\left(\Leftrightarrow y_t=\prod_n{I_n^t}\right)$.

%Two equations (\ref{ineq}) and (\ref{nyan})
%display the action of time evolution
%to the group $\mbox{Pic}^d(C)$ clearly and 
%gives following proposition.
%\begin{prop}
%Let $D_t$ be the divisor 
%\[
%D_t=Q-A,
%\] 
%where $A=(0,y_0)$ 
%and $y_0$ is the complex number which satisfies $\det{R_t(y_0)}=0$.
%Then, the following 
%diagram is commutative.
%\[
%\begin{array}{ccccc}
% &\tee_C& \to &\mbox{Pic}^d(C)& \\[1mm]
%{}_{t\mapsto t+1}& \downarrow& &\downarrow&\hspace{-4mm}{}_{+D_t} \\[1mm]
%  & \tee_C& \to& \mbox{Pic}^d(C)&
%\end{array}
%\]
%\end{prop}

\subsection{The transposed operator}
In this subsection, we introduce the \textit{transposed operator},
and give the proof of proposition \ref{time-shift}.

The section of the eigenvector bundle $V^\lor$ can be described as a rational function
of $x$ and $y$. 
We denote the set of these sections by $\Gamma(V^\lor)$.
Let $\Delta_{i,j}:=(-1)^{i+j}\times (i,j)$-th minor of $X(y)-xE$.
We have
$
g_k=\frac{\Delta_{N,k}}{\Delta_{N,N}} (k=1,2,\dots,N-1)
$,
where $(g_1,\dots,g_{N-1},1)^T\in \Gamma(V^\lor)$. 
We are
interested in the divisor $(g_1)_\infty$ and hence,
it is important
to explore the common zeros of $\Delta_{N,1}$ and $\Delta_{N,N}$.

%We define the involution $\tau:\mbox{Im}\, \varphi_C\to\mbox{Im}\, \varphi_C$ as
%\[
%\begin{array}{ccccccc}
% &\tee_C& \to &\mbox{Im}\, \varphi_C &\subset&\mbox{Pic}^d(C)& \\[1mm]
%{T}\hspace{-4mm}& \downarrow& \circlearrowright&\downarrow&\hspace{-9mm}{\tau} \\[1mm]
%  & \tee_C& \to&\mbox{Im}\, \varphi_C &\subset& \mbox{Pic}^d(C)&
%\end{array},
%\]
%where $T:X(y)\mapsto X(y)^T$. (Note that $X(y)$ and $X(y)^T$ gives the
%same spectral curve $C$.)
Equation (\ref{Lax}) is equivalent to
\begin{equation}\label{tr-Lax}
X_{t}(y)^TR_t(y)^T=R_t(y)^TX_{t+1}(y)^T.
\end{equation}
Let us define $s:=-t$, and $A^\star:=JA^TJ$
for a regular matrix $A$ where
\[
J=\left(\begin{array}{@{\,}cccc@{\,}}
	0 &  &  & 1 \\
	 &  & \mbox{\rotatebox[origin=b]{90}{$\ddots$}} &  \\
	 & 1 &  &  \\
	1 &  &  & 0
\end{array}\right).
\]
Then (\ref{tr-Lax}) becomes 
\begin{equation}\label{gyaku}
X^\star_{s+1}R^\star_{s+1}=R^\star_{s+1}X^\star_{s}.
\end{equation}
Note that the matrix $X^\star(y)$ is also of the form (\ref{matrix}).
We call this new matrix $X^\star$ the \textit{transposed operator}
of $X$. 
A careful analysis of the eigenvector bundle of the transposed operator gives
more information on the original matrix $X$.

\begin{lemma}[van Moerbeke, Mumford]\label{Mumford}\ 
$(1)$\ There exist positive regular divisors $\mathcal{D}_1$, 
$\mathcal{D}_2$, and $\mathcal{D}_3$ such that
\begin{equation}
\textstyle\left(\Delta_{N,1}/\Delta_{N,N}\right)
%\left(\frac{\Delta_{N,1}}{\Delta_{N,N}}\right)
=\mathcal{D}_1+(N-1)P-
\varphi_{\mathrm{fn}}(X)-(N-1)Q,
\end{equation}
and
\begin{equation}\label{tenti}
\textstyle\left(\Delta_{1,N}/\Delta_{N,N}\right)
%\left(\frac{\Delta_{1,N}}{\Delta_{N,N}}\right)
=\mathcal{D}_2+(N-1)Q-\mathcal{D}_3-(N-1)P.
\end{equation}
Moreover, $\deg{\mathcal{D}_i}=g\ (i=1,2,3)$.\\
$(2)$The divisor $(\Delta_{N,N})$ satisfies
\begin{equation}
(\Delta_{N,N})=\varphi_{\mathrm{fn}}(X)+\mathcal{D}_3-(NM-M-n+1)P-NQ.
\end{equation}
\end{lemma}
\begin{rem}
Lemma \ref{Mumford} $(1)$ and Remark \ref{tousiki}
imply that $\deg{\frac{\Delta_{1,N}}{\Delta_{N,N}}}=g+N-1$, and that
$\mathcal{D}_2$ and $\mathcal{D}_3$
do not have common points.
\end{rem}
\begin{lemma}\label{d2}
$\varphi_{\mathrm{fn}}(X^\star)=\varphi_{\mathrm{fn}}(X^T)=\mathcal{D}_2$.
\end{lemma}
\proof
Note that $X^\star$ and $X^T$ give the same spectral curve $C$.
By definition, it follows that:
$
X^T(J\vect{v})=x(J\vect{v})\ \Leftrightarrow\ 
X^\star\vect{v}=x\vect{v}
$,
which implies the first equality of the lemma.\\
%Let us recall the correspondence of two sets of
%data $X$ and $\{C,P,Q\}$ displayed in theorem \ref{oneone}.
%To finish the proof, we assume the following lemma \ref{tenten}.
%If we admit this lemma,
The second equality is obtained from
(\ref{jyuuyou}), (\ref{tenti}) and the fact that
the $(N,1)$-th minor of $X$ is the $(1,N)$-th minor of $X^T$.$\qed$
%\begin{lemma}\label{tenten}
%When 
%\[
%X\stackrel{\mbox{1:1}}{\leftrightarrow}\{C,P,Q\}
%\]
%is a correspondence of two sets of data in theorem \ref{oneone},
%then 
%\[
%X^T\stackrel{\mbox{1:1}}{\leftrightarrow}\{C,Q,P\}.
%\]
%\end{lemma}
%\proof []\\
%The equation (\ref{tenti}) and lemma \ref{d2} give the following 
%proposition.

Lemma \ref{d2} and (\ref{tenti}) yield
\begin{eqnarray*}
\mathcal{D}_3\,\sim\,\varphi_{\mathrm{fn}}(X^\star)+(N-1)Q-(N-1)P
\,\sim\,\varphi_{\mathrm{fn}}(X^\star)-Q+P.
\end{eqnarray*}
Using proposition \ref{index-shift}, we obtain
$
\mathcal{D}_3\sim\varphi_{\mathrm{fn}}(\sigma(X^\star))=
\varphi_{\mathrm{fn}}((\sigma^{-1}X)^\star).
$
In fact, by virtue of the
Riemann-Roch theorem, we obtain the following stronger result:
\[
\mathcal{D}_3=\varphi_{\mathrm{fn}}((\sigma^{-1}X)^\star)
\]
because $\mathcal{D}_3$ is a regular divisor of degree $g$.
\begin{prop}\label{zeros}
$\Delta_{N,N}$ has $2g$ zeros in $C\setminus\{x=\infty\}$.
And the corresponding divisor is 
\[
\varphi_{\mathrm{fn}}(X)+\varphi_{\mathrm{fn}}((\sigma^{-1}X)^\star).
\]
\end{prop}
As a corollary of this proposition, the zeros of $\Delta_{1,1},\Delta_{1,N},\Delta_{N,1}$
are also determined:
\begin{cor}\label{zeros2}
$\Delta_{1,1},\Delta_{1,N},\Delta_{N,1}$ have $2g$ zeros in $C\setminus\{x=\infty\}$
respectively. The corresponding divisors are:
\begin{equation} 
\begin{array}{lcrcl}
\Delta_{1,1}&:&\varphi_{\mathrm{fn}}(\sigma^{-1}X)&+&\varphi_{\mathrm{fn}}(X^\star)\\
\Delta_{1,N}&:&\varphi_{\mathrm{fn}}(X)&+&\varphi_{\mathrm{fn}}(X^\star)\\
\Delta_{N,1}&:&\varphi_{\mathrm{fn}}(\sigma^{-1} X)&+&\varphi_{\mathrm{fn}}
((\sigma^{-1}X)^\star).
\end{array}
\end{equation}
\end{cor}
\proof
The divisor of $\Delta_{1,1}$ is obtained from $\Delta_{N,N}$ by the change of the index:
$n\to n-1$.
%\[\Delta_{N,N}\stackrel{n\to n+1}{\longmapsto}\Delta_{1,1}.\]
The remaining divisors are determined by the identity
$
\Delta_{1,1}\Delta_{N,N}=\Delta_{1,N}\Delta_{N,1}
$
and (\ref{tousiki}).$\hfill\blacksquare$
%Let us define the set $\mathcal{A}:=\{\alpha_1^{(1)},\dots,\alpha_N^{(1)};
%\dots;\alpha_1^{(M)},\dots,\alpha_N^{(M)};\beta_1,\dots,\beta_N\}$
%and the map $\tau:\mathcal{A}\to\mathcal{A}$ which
%is associated with the correspondence $X\to (\sigma^{-1}X)^\star$,
%as in the example below. 
%Note that $\tau^2=id$.
\begin{example}\label{iti}$($$N=4,M=2$$)$
\[{} X=
\left(\begin{array}{@{\,}cccc@{\,}}
	\alpha_1^{(1)} & \alpha_2^{(2)} & 1 & \beta_4 1/y \\
	\beta_1 & \alpha_2^{(1)} & \alpha_3^{(2)} & 1 \\
	y & \beta_2 & \alpha_3^{(1)} & \alpha_4^{(2)} \\
	\alpha_1^{(2)}y & y & \beta_3 & \alpha_4^{(1)}
\end{array}\right),
X^\star=
\left(\begin{array}{@{\,}cccc@{\,}}
	\alpha_4^{(1)} & \alpha_4^{(2)} & 1 & \beta_4 1/y \\
	\beta_3 & \alpha_3^{(1)} & \alpha_3^{(2)} & 1 \\
	y & \beta_2 & \alpha_2^{(1)} & \alpha_2^{(2)} \\
	\alpha_1^{(2)}y & y & \beta_1 & \alpha_1^{(1)}
\end{array}\right),
\]
\[{} \sigma^{-1}X\!=\!
\left(\begin{array}{@{\!}cccc@{\!}}
	\alpha_4^{(1)} & \alpha_1^{(2)} & 1 & \beta_3 1/y \\
	\beta_4 & \alpha_1^{(1)} & \alpha_2^{(2)} & 1 \\
	y & \beta_1 & \alpha_2^{(1)} & \alpha_3^{(2)} \\
	\alpha_4^{(2)}y & y & \beta_2 & \alpha_3^{(1)}
\end{array}\right),
(\sigma^{-1}X)^\star\!\!=\!\!
\left(\begin{array}{@{\!}cccc@{\!}}
	\alpha_3^{(1)} & \alpha_3^{(2)} & 1 & \beta_3 1/y \\
	\beta_2 & \alpha_2^{(1)} & \alpha_2^{(2)} & 1 \\
	y & \beta_1 & \alpha_1^{(1)} & \alpha_1^{(2)} \\
	\alpha_4^{(2)}y & y & \beta_4 & \alpha_4^{(1)}
\end{array}\right).
\]
The correspondence $X\leftrightarrow (\sigma^{-1}X)^\star$ is equivalent to 
the correspondence:
\[
\alpha_i^{(1)}\leftrightarrow\alpha_{4-i}^{(1)},\ \ 
\alpha_i^{(2)}\leftrightarrow\alpha_{1-i}^{(2)},\ \ 
\beta_i\leftrightarrow\beta_{3-i}.
\]
\end{example}

Our aim is to analyze
the distribution of common zeros of $\Delta_{N,1}$ and $\Delta_{N,N}$.
This set of zeros has the following
characteristic property.
\begin{lemma}\label{commonzeros}
\begin{align*}
&\{\mbox{The common zeros of }\{\Delta_{N,1},\Delta_{N,N}\}\}\\
&{}=\{ \mbox{The common zeros of }\{\Delta_{N,k}\}_{k=1,2,\dots,N}\}.
\end{align*}
\end{lemma}
\proof Consider the following open subset on $C$:
$\widetilde{C}=C\setminus \{x=\infty\}$.
By (\ref{futousiki}) and (\ref{tousiki}), for $k\geq 2$, we obtain
%\begin{equation}
$(g_k)_\infty\vert_{\widetilde{C}}< (g_1)_\infty\vert_{\widetilde{C}}$.
%\end{equation}
This is equivalent to
$
%\begin{equation}
\left(\Delta_{N,k}/\Delta_{N,N}\right)_\infty
\!\!\!<\ \left(\Delta_{N,1}/\Delta_{N,N}\right)_\infty,
%\end{equation}
$
which proves the lemma.$\hfill\blacksquare$

Using the preceding calculations, we come to the linearization result.
\begin{prop}\label{time-shift}
Let $D_{(j)}$ $(j=1,2,\dots,M)$ be the divisors 
$
D_{(j)}=A_j-Q,
$ 
where $A_j=(0,y_j)$ 
and $y_j$ is the complex number which satisfies $y_j=\prod_n{I_n^{j}}$.
If $t\equiv j\pmod{M}$, the following 
diagram is commutative.
\[
\begin{array}{ccccc}
 &\tee_C& \to &\mbox{Pic}^d(C)& \\[1mm]
{}_{t\mapsto t+1}& \downarrow& &\downarrow&\hspace{-4mm}{}_{+D_{(j)}} \\[1mm]
  & \tee_C& \to& \mbox{Pic}^d(C)&
\end{array}
\]
\end{prop}
%We can now proceed with the proof of proposition \ref{time-shift}.
\proof
Recall that $\prod_n{I_n^t}=\prod_n{I_n^{t+M}}$ by lemma \ref{new-lemma}.
Let $t\equiv j\pmod{M}$.
Proposition \ref{zeros} and corollary \ref{zeros2} imply
\[
\{\mbox{Common zeros of }\Delta_{N,1}\mbox{ and }\Delta_{N,N}\}=\varphi_{\mathrm{fn}}
((\sigma^{-1}X)^\star).
\]
Using this fact, it becomes clear that the equations (\ref{ineq}) and (\ref{nyan})
give the time evolution action of the transposed operator $X^\star$.
In fact, by virtue of lemma \ref{commonzeros},
(\ref{ineq}) requires the existence of a point $(x,y)=(0,y')$
which becomes the common zero of $\Delta_{N,1}$ and $\Delta_{N,N}$
when $t\mapsto t+1$. Then (\ref{nyan}) also requires
the existence of a point $(x,y)=(x',y_j)$
which becomes the common zero of $\Delta_{N,1}$ and $\Delta_{N,N}$.
Using the same argument about poles, it follows that the
divisor
$
\mathcal{D}'_{(j)}=Q-A_j
$
obeys the %following
relation:
$
%\begin{equation}
\varphi_{\mathrm{fn}}(X^\star_{t+1})=\varphi_{\mathrm{fn}}(X^\star_t)+\mathcal{D'}_{(j)}.
%\end{equation}
$
Recalling (\ref{gyaku}), the two time evolutions defined by $X_t$
and $X^\star_s$ are opposite: $s=-t$.
Hence the divisor $\mathcal{D}_{(j)}=-\mathcal{D}'_{(j)}=A_j-Q$
gives the time evolution of original system.$\hfill\blacksquare$
\section{Theta function solutions}\label{sec3}
\subsection{The distribution of the points $\varphi_{\mathrm{fn}}(X)$}
Let $\varphi_{\mathrm{fn}}(X)=P_1+\cdots+P_g,\ P_j=(x_j,y_j)\in C$.
We are interested in the numbers $x_j\in\CC,\ (j=1,2,\dots,g)$.
Now, consider the resultant of the polynomials \cite{Prasolov}
$\Phi(x,\!y)\!=\!y\det{(X(y)\!-\!xE)}$ and
$\Delta_{N,N}$ as polynomials in $y$.
Let us denote this resultant by $\mbox{Rest}_y(\Phi,\Delta_{N,N})=:R(x)$,
which is \phantom{t}a\phantom{h} polynomial in $x$.
More precisely, $R(x)$ is an element of $\CC
[\alpha_1^{(1)},\dots,\alpha_N^{(1)};
%\alpha_1^{(2)},\dots,\alpha_N^{(2)};
\dots;\alpha_1^{(M)},\dots,\alpha_N^{(M)};\beta_1,\dots,\beta_N][x]$.
From proposition \ref{zeros}, it follows that $\deg_x{R(x)}=2g$.

We also consider the resultant
$S(x):=\mbox{Rest}_y(\Phi,\Delta_{1,N})$.
Due to proposition \ref{zeros} and corollary \ref{zeros2}
the common divisor of $R(x)$ and $S(x)$ is
an element of 
$\CC(\alpha_1^{(1)},\dots,\alpha_N^{(1)};
\dots;\alpha_1^{(M)},\dots,\alpha_N^{(M)};\beta_1,\dots,\beta_N)[x]$,
the degree of which (as a polynomial in $x$) is equal to $g$.
Multiplying divisors (if needed), we obtain the monic polynomial
\[
\Upsilon(x)\in\CC(\alpha_1^{(1)},\dots,\alpha_N^{(1)};
\dots;\alpha_1^{(M)},\dots,\alpha_N^{(M)};\beta_1,\dots,\beta_N)[x],
\]
the roots of which are the common roots of $R(x)$ and $S(x)$,
i.e., the common roots of $\Delta_{1,N}$ and $\Delta_{N,N}$.
%\begin{prop}\label{bunkai}
%There exist two elements \[f_1(x),f_2(x)
%\in\CC[\alpha_1^{(1)},\dots,\alpha_N^{(1)};
%\dots;\alpha_1^{(M)},\dots,\alpha_N^{(M)};\beta_1,\dots,\beta_N][x]\]
%such that $R(x)=f_1(x)f_2(x)$ and $f_1^\tau(x)=f_2(x)$.\\
%$($A polynomial $f^\tau(x)$ is defined by:
%\[
%f(x)=\sum_i{a_ix^i}\mapsto f^\tau(x)=\sum_i{\tau(a_i)x^i},\quad
%\mbox{where}\ \  a_i\in\mathcal{A}.)\]
%\end{prop}
%The proof of this proposition is purely algebraic, and will be given
%in Appendix.
%
%We denote the ring $\CC[\alpha_1^{(1)},\dots,\alpha_N^{(1)};
%\dots;\alpha_1^{(M)},\dots,\alpha_N^{(M)};\beta_1,\dots,\beta_N]$
%by $\CC[\mathcal{A}]$ for short.
Recalling that the set of the common zeros of $\Delta_{N,N}$ and $\Delta_{1,N}$
contained in $C\setminus \{x=\infty\}$ is $\{P_1,\dots,P_g\}$,
we conclude:
\begin{equation}\label{upsi}
\deg_x{\Upsilon(x)}=g,\quad\
\Upsilon(x_i)=0,\quad\!\! i=1,2,\dots,g.
\end{equation}
\subsection{Theta function solutions}
For a complex curve (or Riemann surface)
$C$ of genus $g$, one usually considers
a \textit{canonical basis of} $H_1(C,\ZZ)$.
We denote the canonical basis by $a_1,\dots,a_g;$
$b_1,\dots,b_g
\in H_1(C,\ZZ)$. Let $\omega_1,\dots,\omega_g$ be the holomorphic differential
of $C$ which satisfies $\int_{a_j}{\omega_i}=\delta_{j,i}$.
A \textit{period matrix} of the Riemann surface $C$ 
is a $g\times g$ matrix $B=(\int_{b_j}\omega_i)$.
Let $\theta(\vect{z},B)$ be the \textit{theta function}; $\CC^g\to\CC$,
and 
$\vect{A}:\mbox{Pic}^g\stackrel{\sim}{\to} J(C)\,(:\,=\CC^g/(\ZZ^g+B\ZZ^g))$
the \textit{Abelian mapping}.
The following theorem is a classical and fundamental result.
\begin{thm}[Riemann]
Let $C$ be a Riemann surface of genus $g$, and let $\mathcal{D}=P_1+\dots+P_g$
be a regular positive divisor. Then the function 
\[F(p)=\theta(\vect{A}(p)-\vect{A}(\mathcal{D})-\vect{K},B),\quad
p\in C\]
has exactly $g$ zeros $p=P_1,\dots,P_g$ on $C$,
where $\vect{K}$ is the Riemann constant of $C$.
\end{thm}
To obtain the solution to the pd Toda equation, we consider the following integral:
\begin{equation}\label{integral}
I=\frac{1}{2\pi i}\int_{\partial C^\circ}{\!\!x(p)\frac{dF(p)}{F(p)}}
\left(=:\frac{1}{2\pi i}\int_{\partial C^\circ}{d\mathcal{Z}(p)}\right).
\end{equation}
Here the integral path $\partial C^\circ$ goes along the edge
of the simply connected domain $C^\circ$ obtained from the Riemann surface
by cutting it along $a_1,\dots,a_g;b_1,\dots,b_g$.

The integral $I$ can be rewritten as
\begin{eqnarray*}
&{} \textstyle
I=\frac{1}{2\pi i}\sum_{k=1}^g\left(\int_{a_k}+\int_{a_k^{-1}}
+\int_{b_k}+\int_{b_k^{-1}}
\right)d\mathcal{Z}(p)\\
&\textstyle{}
=\frac{1}{2\pi i}\sum_{k=1}^g\left(\int_{a_k}{\{d\mathcal{Z}(p)-
d\mathcal{Z}(p+b_k)\}}
+\int_{b_k}{\{d\mathcal{Z}(p)-
d\mathcal{Z}(p+a_k)\}}
\right).
\end{eqnarray*}
Recalling the classical fact
$
%\begin{eqnarray}
%&
F(p+a_k)=F(p),%\\
%&
F(p+b_k)=\exp(-2\pi i(\vect{A}(p)-\vect{A}(\mathcal{D})-\vect{K})_k)F(p)
%\end{eqnarray}
$
and
$
%\begin{equation}
d(\vect{A}(p))_k=\omega_k(p)
%\end{equation}
$,
the integral $I$ is transformed to
\begin{equation}\label{ai1}
\textstyle
I=\sum_{k=1}^g{\int_{a_k}{\!\!x(p)\,\omega_k(p)}}.
\end{equation}
On the other hand, by the residue theorem, the integral $I$ also has the
expression:
\begin{equation}\label{ai2}
\textstyle
I=\sum_{i=1}^g{x(P_i)}+\mbox{Res}_P(d\mathcal{Z})+\mbox{Res}_Q(d\mathcal{Z}).
\end{equation}
Let $t_P$ and $t_Q$ be local coordinates around $P$ and $Q$
respectively.
These satisfy
\[
x\sim 1/(t_P)^M,\,(\mbox{neighbor of }P),\quad
x\sim 1/t_Q,\,(\mbox{neighbor of }Q).
\]
In a neighborhood of $P$, one has:
\begin{eqnarray}
d\mathcal{Z}\,\sim\,
\frac{1}{{(t_P)}^M}\frac{d\log{F}}{dt_P}\,dt_P
\,=\,\frac{1}{(t_P)^{M}}\sum_{l=1}^g{(\partial_l\log{F})
\left(\frac{d(\vect{A})_l}{dt_P}\right)}\,dt_P\label{residue}
\end{eqnarray}
%Let $\Upsilon(x)=x^g-a_1x^{g+1}+\cdots$ be a polynomial in the previous section.
%\begin{thm}
%\[
%\frac{a_1}{a_0}=
%\]
%\end{thm}
To calculate the residue of the differential (\ref{residue}),
we explore the behaviour of 
$
\frac{d(\vect{A})_l}{dt_P}dt_P=\omega_l
$
around points $P$ and $Q$.
Let $c_l:=\mbox{Res}_P(\omega_l/(t_P)^M)$. Then we obtain the expression:
\begin{equation}
\textstyle
\mbox{Res}_P(d\mathcal{Z})=\sum_{l=1}^g{c_l(\partial_l \log(F(P)))}.
\end{equation}
In the similar manner, we also conclude
\begin{equation}
\textstyle
\mbox{Res}_Q(d\mathcal{Z})=\sum_{l=1}^g{c'_l(\partial_l \log(F(Q)))},
\end{equation}
where $c'_l:=\mbox{Res}_Q(\omega_l/t_Q)$.

By (\ref{ai1}) and (\ref{ai2}), we obtain
\begin{equation}
{}
\sum_{l=1}^g{x(P_l)}=\sum_{l=1}^g{\int_{a_l}{\!\!x(p)\,\omega_l(p)}}
-\sum_{l=1}^g{c_l(\partial_l \log(F(P)))}-\sum_{l=1}^g{c'_l(\partial_l \log(F(Q)))}.
\end{equation}
Using this equation, we obtain the following theorem which is a generalization
of the preceding result concerning the
theta function solution to the pd Toda $(M=1)$ equation \cite{Kimijima}
and which is 
the main theorem in the present paper:
\begin{thm}\label{seoremu}
Let $\Upsilon(x)$ be the monic polynomial of degree $g$ obtained by
$(\mathrm{\ref{upsi}})$:
\[
\Upsilon(x)=x^g-a_1x^{g-1}+\dots+(-1)^ga_g,
\]
with $a_1,\dots,a_g\in\CC(\alpha_1^{(1)},\dots,\alpha_N^{(1)};
\dots;\alpha_1^{(M)},\dots,\alpha_N^{(M)};\beta_1,\dots,\beta_N)$.
Then we find
\begin{eqnarray}
a_1&=\sum_{l=1}^g{\int_{a_l}{\!\!x(p)\,\omega_l(p)}}
-\sum_{l=1}^g{c_l\,\{\partial_l \log{\theta(n\vect{k}+
\vect{\nu}(t)+\vect{c}_0,B)}\}}\nonumber\\
&\ \ \ \ -\sum_{l=1}^g{c'_l\,\{\partial_l \log{\theta((n+1)\vect{k}+\vect{\nu}(t)
+\vect{c}_0,B)}\}},
\end{eqnarray}
where $
c_l=\mbox{Res}_P(\omega_l/(t_P)^M)$, 
$c'_l=\mbox{Res}_Q(\omega_l/t_Q)$,
$\vect{k}=\vect{A}(P-Q)$, $\vect{c}_0=\vect{A}(Q-\mathcal{D}_0+\Theta)$, and
$
\vect{\nu}(pM+q)=p\vect{A}(A_1+\dots+A_M)
+\vect{A}(A_1+\dots+A_q)-(pM+q)\vect{A}(Q)
$ $(1\leq q\leq M)$.
Here $\Theta$ is the theta divisor: $\vect{A}(\Theta)=-\vect{K}$.
The divisor $\mathcal{D}_0$ is the initial value
$
\varphi_{\mathrm{fn}}(X_{t=0}(y))=\mathcal{D}_0
$.
\end{thm}
\subsection*{Acknowledgement}
The author is very grateful to Professor Teshuji Tokihiro and
Professor Ralph Willox for helpful comments 
on this paper. 
\appendix
\section{Proof of lemma \ref{det}}
\begin{figure}[htbp]
\begin{center}
\begin{picture}(-100,120)(50,-120)
\put(0,0){$1$}
%-----------------------
\put(-40,-40){$I_1^t$}
\put(40,-40){$1$}
%--------------------------------
\put(-90,-80){$I_1^tI_1^{t+1}$}
\put(-10,-80){$I_2^t+I_1^{t+1}$}
\put(80,-80){$1$}
%--------------------------------
\put(-160,-120){$I_1^tI_1^{t+1}I_1^{t+2}$}
\put(-90,-120){$I_2^{t}I_2^{t+1}+I_2^tI_1^{t+2}+I_1^{t+1}I_1^{t+2}$}
\put(40,-120){$I_3^t+I_2^{t+1}+I_1^{t+2}$}
\put(120,-120){$1$}
%-----------------------------------------
\put(8,-4){\vector(1,-1){30}}
\put(48,-44){\vector(1,-1){30}}
\put(88,-84){\vector(1,-1){30}}
\put(128,-124){\vector(1,-1){10}}
\put(-32,-44){\vector(1,-1){27}}
\put(18,-84){\vector(1,-1){27}}
\put(58,-124){\vector(1,-1){10}}
\put(-75,-84){\vector(1,-1){27}}
\put(-35,-124){\vector(1,-1){10}}
\put(-125,-124){\vector(1,-1){10}}
%--------------------------------------------
\put(-2,-4){\vector(-1,-1){27}}
\put(38,-44){\vector(-1,-1){30}}
\put(78,-84){\vector(-1,-1){27}}
\put(-42,-44){\vector(-1,-1){25}}
\put(-82,-84){\vector(-1,-1){30}}
\put(-2,-84){\vector(-1,-1){25}}
\put(-130,-124){\vector(-1,-1){10}}
\put(-40,-124){\vector(-1,-1){10}}
\put(55,-124){\vector(-1,-1){10}}
\put(120,-124){\vector(-1,-1){10}}
%------------------------------------
\put(-40,-15){$\times I_1^t$}
\put(-90,-55){$\times I_1^{t+1}$}
\put(5,-55){$\times I_1^{t+1}$}
\put(-130,-95){$\times I_1^{t+2}$}
\put(-40,-95){$\times I_1^{t+2}$}
\put(45,-95){$\times I_1^{t+2}$}
%-------------------------------------
\put(40,-15){$\sigma$}
\put(80,-55){$\sigma$}
\put(120,-95){$\sigma$}
\put(-16,-55){$\sigma$}
\end{picture}
\end{center}
\caption[Sumation]{The $j+1$-th row of this diagram displays
the first $j+1$ non-zero
components of the row vector $\vect{u}^{(j)}$.}
\label{sumation}
\end{figure}
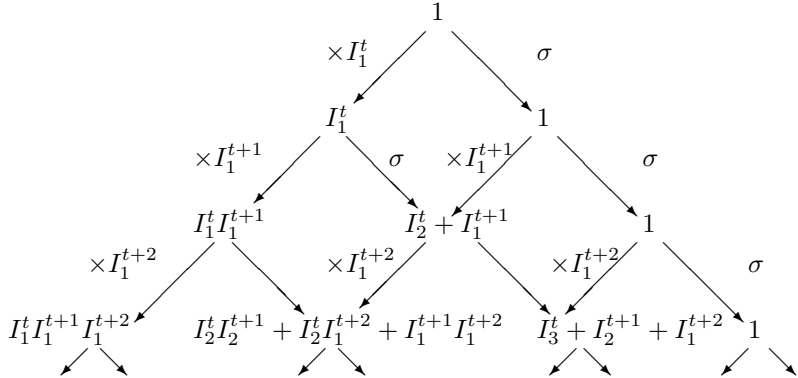

In this section, we give the proof of lemma \ref{det}.
Let \[U_{k}:=R_{t+k-1}(y)R_{t+k-2}(y)\dots R_t(y),\quad
1\leq k\leq M.\]
Define the row vector
$\vect{u}^{(k)}=(u_{1,1}^{(k)},u_{1,2}^{(k)},\dots,u_{1,N}^{(k)})$,
where $U_k=(u_{i,j}^{(k)})$, and
the following homomorphism of rings:
\[
\sigma:\ZZ\,[\{V_n^t,I_n^t\}_{n\in\ZZ}]\to
\ZZ\,[\{V_n^t,I_n^t\}_{n\in\ZZ}]\quad;\quad
V_n^t\mapsto V_{n+1}^t,\,I_n^t\mapsto I_{n+1}^t.
\]
We rewrite
the $j$-th component of $\vect{u}^{(k)}$ as $u_j^{(k)}$ for short.
By definition of $R_t(y)$, we obtain
\begin{equation}\label{algo}
u_{j}^{(k+1)}=\sigma(u_{j-1}^{(k)})+I_1^{t+k}u_{j}^{(k)},\quad (u^{(k)}_{-1}=0).
\end{equation}
On the other hand, the $2$nd row of the matrix $X_t(y)$,
which is of the form
\[(\beta_1,\alpha_2^{(1)},\alpha_3^{(2)},\dots,\alpha_{M+1}^M,1,0,\dots,0),\]
satisfies
$
%\begin{equation}
\beta_1=V_1^tu_1^{(M)},\ \alpha_{j+1}^{(j)}=\sigma(u_{j}^{(M)})+V_{1}^tu_{j+1}^{(M)}.
%\end{equation}
$
Figure \ref{sumation} displays the algorithm which we shall use
to obtain the row vector $\vect{u}^{(k)}$ expressed by (\ref{algo}).

Let us introduce the signs $\swarrow$ and $\searrow$
to describe the terms which appear in the figure \ref{sumation}.
Let us define the set of arrows $Ar:=\{\swarrow,\searrow\}$
and the set of sequences $Ar^{r}:=
\{(a_1,a_2,\dots,a_r)\,\vert\, a_j\in Ar\},\,r\in\NN$.
We define the map of sets $\{\cdot\}:
\bigcup_r{Ar^{r}}\to \ZZ[\{I_n^{t+l}\}_{1\leq n\leq N,1\leq l\leq M}]$
as follows:\\
For $r=0$, define $\{\emptyset\}:=1$.
If $\{a_1,a_2,\dots,a_r\}\in \ZZ[\{I_n^{t+l}\}_{n,l}]$ is given,
we define
$\{a_1,a_2,\dots,a_r,a_{r+1}\}$
by
\[{}
\{a_1,\dots,a_r,\swarrow\}:=I_1^{t+r}\{a_1,a_2,\dots,a_r\},\quad
\{a_1,\dots,a_r,\searrow\}:=\sigma(\{a_1,\dots,a_r\})
\]
inductively.
For example,
$
\{\swarrow\}=I_1^t$, $
\{\swarrow\swarrow\}=I_1^tI_1^{t+1}$, $
\{\swarrow\searrow\}=I_2^t$, $
\{\searrow\swarrow\}=I_1^{t+1}.
$
By definition, the $j$-th component of $\vect{u}^{(k)}$ satisfies
\begin{equation}
u_j^{(k)}=\sum_{\sharp\searrow=j-1,\
\sharp\swarrow=k-j+1}{\!\!\!\!\!\!\!\{a_1,\dots,a_k\}}.
\end{equation}
\begin{lemma}\label{lemmaa}
$
\{\swarrow,a_2,\dots,a_k\}=I_{l+1}^{t}\{\searrow,a_2\dots,a_k\},
$\\
where $l=\sharp\{\mbox{the arrow $\searrow$ included in $\{a_2,\dots,a_k\}$}\}$.
\end{lemma}
\proof 
We prove the equation by induction respected to $k$.
If $k=1$, the equation is equivalent to $\{\swarrow\}=I_1^t\{\searrow\}$ which
is true by definition.
Let $k\geq 2$. If $a_{k}=\swarrow$, we obtain
$\mbox{l.h.s}=I_1^{t+k-1}\{\swarrow,a_2,\dots,a_{k-1}\}$,
$\mbox{r.h.s}=I_1^{t+k-1}I_{l+1}^t\{\searrow,a_2,\dots,a_{k-1}\}$.
By assumption of induction, it follows that $\mbox{l.h.s}=\mbox{r.h.s}$.
We also prove the equation in the similar manner
if $a_{k}=\searrow$. 
$\qed$
\begin{lemma}\label{lemmab}
$
u_1^{(M)}+\sum_{j=1}^M{(-1)^ju_{j+1}^{(M)}I_1^t\cdots I_j^t}=0.
$
\end{lemma}
\proof 
\begin{align*}
&\textstyle
u_1^{(M)}+\sum_{j=1}^M{(-1)^ju_{j+1}^{(M)}I^t_1\cdots I^t_j}\\
&{}=\{\swarrow\swarrow\cdots\swarrow\}+\sum_{j=1}^M
{(-1)^j\left[\sum_{\sharp\searrow=j,\ \sharp\swarrow=M-j}
{\hspace{-0.8cm}
\{\swarrow,\ast,\dots,\ast\}+\{\searrow,\ast,\dots,\ast\}}\right]I^t_1\cdots I^t_j}\\
&=\{\swarrow\swarrow\cdots\swarrow\}+
\sum_{j=1}^{M-1}{(-1)^j\hspace{-0.5cm}\sum_{\sharp\searrow=j+1,\,\sharp\swarrow=M-j-1}
{\hspace{-1cm}\{\searrow,\ast\}\,I^t_1\cdots I_j^tI^t_{j+1}}}\\
&+\sum_{j=1}^{M}{(-1)^j\hspace{-0.5cm}\sum_{\sharp\searrow=j,\,\sharp\swarrow=M-j}
{\hspace{-0.7cm}
\{\searrow,\ast\}\,I^t_1\cdots I^t_j}}
\quad\quad(\because \mbox{Lemma \ref{lemmaa}})\\
&=\{\swarrow\swarrow\cdots\swarrow\}-I_1\{\searrow\swarrow
\cdots\swarrow\}\\
&=0 \quad\quad\quad\quad
\qed
\end{align*}
\begin{lemma}\label{lemmac}
$
\sum_{j=1}^M{(-1)^j\sigma(u_j^{(M)})I_1^t\cdots I_j^{t}}=(-1)^MI_1^t\cdots I_{M+1}^t.
$
\end{lemma}
\proof We start from lemma \ref{lemmab}.
\begin{align*}
\textstyle
0&=\textstyle\sigma\left(u_1^{(M)}+\sum_{j=1}^M{(-1)^ju_{j+1}^{(M)}I^t_1\cdots I^t_j}
\right)\\
&=\textstyle\sigma(u_1^{(M)})+\sum_{j=1}^M{(-1)^j\sigma(u_{j+1}^{(M)})I^t_2\cdots I^t_{j+1}}.
\end{align*}
Multiplying $I_1$, we obtain
\begin{align*}
0&=\textstyle\sum_{j=1}^{M+1}{(-1)^j\sigma(u_j^{(M)})I^t_1\cdots I^t_j}\\
&=\textstyle
\sum_{j=1}^M{(-1)^j\sigma(u_j^{(M)})I^t_1\cdots I^t_j}+(-1)^{M+1}I^t_1\cdots I^t_{M+1},
\end{align*}
which complete the proof. $\qed$

\underline{proof of lemma \ref{det}}\\
Calculate $\det{H_t}$ by the definition of $H_t$
(\ref{eiti}). The components $v_{M+2}^{(j)},\ (j=1,2,\dots,M+1)$ are rewritten as
\[{}
v_{M+2}^{(1)}=-\beta_1,\ v_{M+2}^{(2)}=x-\alpha_2^{(1)},
\ v_{M+2}^{(j)}=-\alpha_j^{(j-1)},\,(j=3,4,\dots,M+1).
\]
by virtue of (\ref{linear}).
The expansion of the determinant with respect to the $(M+1)$-st row yields
\begin{align*}
\det{H_t}
&
=(-1)^M\{
-\beta_1+(\alpha_2^{(1)}-x)I_1^t-\alpha_3^{(2)}I_1^tI_2^t+
\cdots\\
&\hspace{4cm}%
+(-1)^{M}(I_{M+1}^t-\alpha_{M+1}^{(M)})
I_1^t\cdots I_{M}^t
\}\\
&%\ \ \ \ \ \ \ \ \ \ {}
\textstyle
=\!\!(-1)^M\left\{-\beta_1\!-\!xI_1^t-\sum_{j=1}^M{(-1)^j\alpha_{j+1}^{(j)}I_1^t\cdots I_{j}^t}
+(-1)^MI_1^t\cdots I_M^tI_{M+1}^t\right\}\\
&%\ \ \ \ \ \ \ \ \ \ {}
\textstyle
=(-1)^M\left\{-V_1^tu_1^{(M)}-xI_1^t-\sum_{j=1}^M{(-1)^j\{\sigma(u_j^{(M)})
+V_1^tu_{j+1}^{(M)}
\}I_1^t\cdots I_{j}^t}\right.\\
&%\ \ \ \ \ \ \ \ \ \ {}
\hspace{6cm}\textstyle\left.
\phantom{\sum_{j}{lo}}+(-1)^MI_1^t\cdots I_M^tI_{M+1}^t\right\}\\
&%\ \ \ \ \ \ \ \ \ \ {}
=(-1)^{M+1}I_1^t x. \quad(\because
\mbox{Lemma \ref{lemmab} and \ref{lemmac}})\quad\quad\quad\qed
\end{align*}
%\section*{References}

\end{document}